\documentclass[twocolumn,pra,aps,showpacs,superscriptaddress,floatfix,showkeys,nofootinbib,10pt]{revtex4-2}

\usepackage[T1]{fontenc}
\usepackage{amsmath,amsfonts,amssymb,amsthm}
\usepackage{graphicx}

\usepackage{tikz}
\usetikzlibrary{arrows.meta, decorations.pathmorphing, positioning, shapes.misc, calc, fit}

\newcommand{\ket}[1]{ | \, #1 \rangle} \newcommand{\bra}[1]{ \langle #1 \, |} 
\newcommand{\proj}[1]{\ket{#1}\bra{#1}}

\newcommand{\be}{\begin{equation}} \newcommand{\ee}{\end{equation}}
\newcommand{\ba}{\begin{aligned}} \newcommand{\ea}{\end{aligned}}

\DeclareMathOperator{\Tr}{Tr}

\DeclareMathOperator{\Cor}{Cor}

\DeclareRobustCommand\openone{\leavevmode\hbox{\small1\normalsize\kern-.33em1}}%

\begin{document}

\title{When More Light Means Less Quantum: Modeling Bell Inequality Degradation from Accidental Counts}
\author{Piotr Mironowicz} \email{piotr.mironowicz@gmail.com}
\affiliation{Department of Algorithms and System Modeling, Faculty of Electronics, Telecommunications and Informatics, Gda\'{n}sk University of Technology, Poland}
\affiliation{Center for Theoretical Physics, Polish Academy of Sciences, Aleja Lotników 32/46, 02-668 Warsaw, Poland}
\author{Mohamed Bourennane}
\affiliation{Department of Physics, Stockholm University, S-10691 Stockholm, Sweden}

\date{\today}

\begin{abstract}
	We investigate how accidental counts, the detection events not originating from genuine entangled photon pairs, impact the observed violation of Bell inequalities in photonic experiments. These false coincidences become increasingly significant at higher laser pump powers, limiting the strength of Bell violations and thus the performance of quantum protocols such as device-independent quantum random number generation and quantum key distribution. We propose a simple noise model that quantitatively links the Bell value to the pump strength. Using experimental data from recent SPDC-based Bell tests, we fit the model to a Bell expression, and demonstrate accurate prediction of Bell values across a range of pump settings. Our results provide practical guidance for optimizing source brightness while preserving quantum nonlocality, with direct implications for high-rate, secure quantum technologies.
\end{abstract}


\maketitle

\section{Introduction}

Bell inequalities are among the most fascinating and counterintuitive results in the foundations of quantum mechanics. Introduced by John Bell in 1964~\cite{bell1964einstein}, they provide a way to test whether the correlations predicted by quantum mechanics and observed in experiments, can be explained by any local hidden variable (LHV) theory. Remarkably, quantum theory predicts correlations that violate these inequalities, a feature confirmed in a series of increasingly precise experiments~\cite{aspect1981experimental,rowe2001experimental,giustina2015significant,shalm2015strong,rosenfeld2017event,hensen2015loophole}. These violations have been interpreted not only as a deep insight into the nature of physical reality but also as practical tools for secure quantum technologies.

In these experiments, the goal is typically to maximize the value of a particular Bell expression, which quantifies the degree of non-classical correlation. There exists a multitude of such expressions beyond the well-known Clauser–Horne–Shimony–Holt (CHSH) inequality~\cite{clauser1969proposed}, including entire families of inequalities optimized for self-testing or for certifying device-independent randomness~\cite{mironowicz2013robustness,brunner2014bell,acin2016certified,kaniewski2019maximal,augusiak2019bell,wooltorton2022tight}. Choosing the right Bell expression is essential in practical applications, as different inequalities respond differently to experimental imperfections and noise.

Modern Bell tests are typically implemented using photonic platforms based on quantum optics, particularly spontaneous parametric down-conversion (SPDC), a process in which a nonlinear crystal converts a high-energy pump photon into a pair of entangled lower-energy photons~\cite{kwiat1995new,burnham1970observation}. The photons are then distributed to spatially separated measurement stations, commonly called Alice and Bob, where their polarization or path is measured.

As the power of the laser pump increases, one naturally expects a higher rate of generated photon pairs, leading to higher event rates and potentially stronger exposition of quantum phenomena. However, this is complicated by the increasing probability of generating multiple pairs simultaneously, which leads to accidental coincidences, i.e. events where uncorrelated photons are mistakenly registered as coincident detections~\cite{ferreira2013proof}. These accidental counts degrade the quality of the observed correlations and can significantly reduce the observed Bell violation. Accurate modeling and mitigation strategies are therefore essential for the success of high-fidelity quantum experiments. Techniques such as narrowing the coincidence window, improving detector timing resolution, and employing heralded single-photon sources are commonly applied to reduce their influence. Statistical subtraction and bounding methods are also widely used in randomness and key rate estimation. Foundational studies like those of Kwiat et al.~\cite{kwiat1995new} brought attention to accidentals in SPDC-based entanglement sources, while modern experimental efforts, including loophole-free Bell tests~\cite{giustina2015significant,bierhorst2018experimentally} and quantum cryptography trials~\cite{ferreira2013proof}, have advanced robust frameworks for quantifying and minimizing accidental counts. Detailed modeling and practical implications are extensively discussed in~\cite{stefanov2000optical, christensen2013detection,liu2018device}. Accidental counts thus remain a fundamental limiting factor in demonstrating quantum advantage in real-world quantum optical systems.

In SPDC sources, the probability of generating $k$ pairs per pump pulse follows a thermal or Poissonian distribution depending on the experimental regime. As the pump power increases, the tail of this distribution becomes non-negligible, increasing the contribution of multi-pair emissions~\cite{christ2011probing,pevrina2013quantum}. Because detectors cannot distinguish whether a detected photon comes from a single-pair or multi-pair event, this leads to noise that is indistinguishable from legitimate entangled events, effectively washing out the quantum correlations~\cite{liu2018high,sekatski2012detector}.

This trade-off between Bell violation and event rate is particularly relevant in device-independent (DI)~\cite{10.5555/2011827.2011830} quantum information protocols such as quantum key distribution (QKD)~\cite{acin2007device,pironio2009device,masanes2011secure} and quantum random number generation (QRNG)~\cite{pironio2010random,mironowicz2013robustness}. In these scenarios, both a strong Bell violation and a high event rate are desired to ensure high-quality randomness or key generation within a reasonable amount of time. Therefore, understanding how the observed Bell value depends on the laser pump power becomes crucial for designing optimal quantum technologies.

In this work, we propose a quantitative model that predicts the value of a Bell expression as a function of the laser pump power. Our model incorporates the probabilistic generation of photon pairs, imperfect detection efficiencies, and the contribution of accidental coincidences. This predictive capability is useful for optimizing experimental setups in QRNG and QKD, where a delicate balance must be struck between maximizing the Bell value and achieving high throughput.

In Subsection~\ref{sec:Bell}, we present the Bell inequalities used as randomness certificates, in particular the well-known CHSH expression, and in Subsection~\ref{sec:SPDC} we discuss SPDC sources. Section~\ref{sec:accidental_counts} discusses the origin and impact of accidental counts in SPDC-based experiments, especially as a function of laser pump power. In Section~\ref{sec:noise_model}, we introduce a noise model that captures the contribution of accidental coincidences to the observed correlations. Section~\ref{sec:predictions} presents a quantitative analysis based on experimental data from~\cite{piveteau2024optimization}, showing how Bell inequality violations degrade with increasing pump strength. Using our noise model, we predict the expected Bell values beyond the range of the original measurements. Finally, in Section~\ref{sec:conclusions}, we summarize our findings and discuss their implications for the design of high-rate, DI quantum technologies such as QRNGs and QKD systems.

\section{Preliminaries}
\label{sec:preliminaries}

\subsection{Bell certificates of randomness}
\label{sec:Bell}

The Tsirelson bound sets the maximum strength of correlations that quantum mechanics allows between measurements on entangled particles~\cite{cirel1980quantum,horodecki2009quantum}. While classical correlations are constrained by a strict limit, the so-called classical bound, quantum mechanics permits stronger correlations that still respect Tsirelson’s limit. These bounds are critical because they serve to differentiate quantum correlations from those explainable through classical mechanisms.

To quantify the correlations between measurement outcomes, we define the correlator function $\Cor(x,y)$, representing the correlation between Alice’s measurement $x$ and Bob’s measurement $y$, as $\Cor(x,y) \equiv P(0,0|x,y) + P(1,1|x,y) - P(0,1|x,y) - P(1,0|x,y)$.
In CHSH scenario, Alice and Bob each select between two possible measurements~\cite{clauser1969proposed}. The CHSH expression is defined by $CHSH \equiv \Cor(0,0)+\Cor(0,1)+\Cor(1,0)-\Cor(1,1)$. The Tsirelson bound of CHSH is $2 \sqrt{2}$, and the classical bound is $2$.


\subsection{Spontaneous parametric down-conversion source of entangled photons}
\label{sec:SPDC}

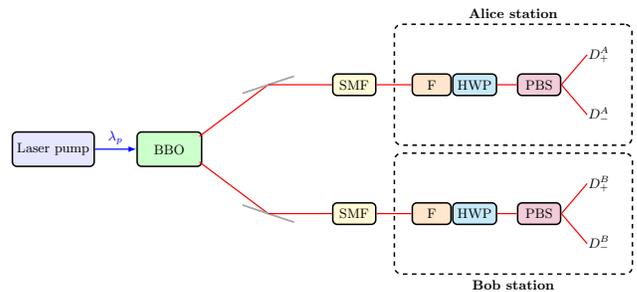
\begin{figure}
    \begin{center}
    	\resizebox{.97\linewidth}{!}{
		    \begin{tikzpicture}[>=latex, line width=1pt]
			    \node[draw, fill=blue!10, minimum width=1.8cm, minimum height=0.8cm, rounded corners=3pt] 
			      (pump) at (0,0) {Laser pump};
			    
			    \node[draw, fill=green!20, minimum width=1.5cm, minimum height=0.8cm, rounded corners=3pt] 
			      (bbo) at (2.7,0) {BBO};
			    
			    \draw[->, blue, thick] (pump) -- (bbo) node[midway, above] {$\lambda_p$};
			    
			    \coordinate (leftPhotonStart) at (3.4,0.3);
			    \coordinate (rightPhotonStart) at (3.4,-0.3);
			    
			    \coordinate (leftMirror) at (5,1.5);
			    \coordinate (rightMirror) at (5,-1.5);
			    
			    \draw[red, thick] (leftPhotonStart) -- (leftMirror);
			    \draw[red, thick] (rightPhotonStart) -- (rightMirror);
			    
			    \draw[gray!70, very thick] ($(leftMirror)+(-0.6,-0.2)$) -- ($(leftMirror)+(0.6,0.2)$);
			    \draw[gray!70, very thick] ($(rightMirror)+(-0.6,0.2)$) -- ($(rightMirror)+(0.6,-0.2)$);
			    
			    \coordinate (leftSMF) at (7,1.5);
			    \coordinate (rightSMF) at (7,-1.5);
			    
			    \draw[red, thick] (leftMirror) -- (leftSMF);
			    \draw[red, thick] (rightMirror) -- (rightSMF);
			    
			    \node[draw, fill=yellow!20, minimum width=1.0cm, minimum height=0.5cm, rounded corners=3pt] 
			      (smfL) at (leftSMF) {SMF};
			    \node[draw, fill=yellow!20, minimum width=1.0cm, minimum height=0.5cm, rounded corners=3pt] 
			      (smfR) at (rightSMF) {SMF};
			    
			    \coordinate (leftFilter) at (8.8,1.5);
			    \coordinate (rightFilter) at (8.8,-1.5);
			    
			    \node[draw, fill=orange!20, minimum width=0.9cm, minimum height=0.5cm, rounded corners=3pt] 
			      (filterL) at (leftFilter) {F};
			    \node[draw, fill=orange!20, minimum width=0.9cm, minimum height=0.5cm, rounded corners=3pt] 
			      (filterR) at (rightFilter) {F};
			    
			    \draw[red, thick] (smfL.east) -- (filterL.west);
			    \draw[red, thick] (smfR.east) -- (filterR.west);
			    
			    \coordinate (aliceStart) at (9.8,1.5);
			    \coordinate (bobStart) at (9.8,-1.5);
			    
			    \node[draw, fill=cyan!20, minimum width=1.0cm, minimum height=0.5cm, rounded corners=3pt] 
			      (hwpA) at (aliceStart) {HWP};
			    \coordinate (pbsPosA) at (11.3,1.5);
			    \node[draw, fill=purple!20, minimum width=1.0cm, minimum height=0.5cm, rounded corners=3pt] 
			      (pbsA) at (pbsPosA) {PBS};
			    
			    \node[draw, fill=cyan!20, minimum width=1.0cm, minimum height=0.5cm, rounded corners=3pt] 
			      (hwpB) at (bobStart) {HWP};
			    \coordinate (pbsPosB) at (11.3,-1.5);
			    \node[draw, fill=purple!20, minimum width=1.0cm, minimum height=0.5cm, rounded corners=3pt] 
			      (pbsB) at (pbsPosB) {PBS};
			    
			    \draw[red, thick] (filterL.east) -- (hwpA.west);
			    \draw[red, thick] (filterR.east) -- (hwpB.west);
			    \draw[red, thick] (hwpA.east) -- (pbsA.west);
			    \draw[red, thick] (hwpB.east) -- (pbsB.west);
			    
			    \draw[red, thick] (pbsA.east) -- ++(0.6,0.7);
			    \draw[red, thick] (pbsA.east) -- ++(0.6,-0.7);
			    
			    \draw[red, thick] (pbsB.east) -- ++(0.6,0.7);
			    \draw[red, thick] (pbsB.east) -- ++(0.6,-0.7);
			    
			    \path (pbsA.east) ++(0.6,0.7) coordinate (detAplus);
			    \path (pbsA.east) ++(0.6,-0.7) coordinate (detAminus);
			    \path (pbsB.east) ++(0.6,0.7) coordinate (detBplus);
			    \path (pbsB.east) ++(0.6,-0.7) coordinate (detBminus);
			    
			    \node[right] (detAplusLabel) at ($(detAplus)+(-0.1,0)$) {$D_{+}^A$};
			    \node[right] (detAminusLabel) at ($(detAminus)+(-0.1,0)$) {$D_{-}^A$};
			    \node[right] (detBplusLabel) at ($(detBplus)+(-0.1,0)$) {$D_{+}^B$};
			    \node[right] (detBminusLabel) at ($(detBminus)+(-0.1,0)$) {$D_{-}^B$};
			    
			    \node[draw, dashed, rounded corners=5pt, inner sep=0.4cm, fit=(filterL)(hwpA)(pbsA)(detAplus)(detAminus)(detAplusLabel)(detAminusLabel)] 
			      (aliceBox) {};
			    \node[draw, dashed, rounded corners=5pt, inner sep=0.4cm, fit=(filterR)(hwpB)(pbsB)(detBplus)(detBminus)(detBplusLabel)(detBminusLabel)] 
			      (bobBox) {};
			    
			    \node[above] at (aliceBox.north) {\textbf{Alice station}};
			    \node[below] at (bobBox.south) {\textbf{Bob station}};
		    \end{tikzpicture}
    	}
    	\end{center}
    \caption{(color online) \textbf{Generic schematic of an SPDC-based Bell experiment.} 
    A pump laser illuminates a nonlinear crystal (e.g., BBO), producing pairs of entangled photons that propagate into two spatial modes. 
    Each photon is directed by mirrors, coupled into single-mode fibers (SMFs), spectrally filtered (F), and sent to a measurement station. 
    Alice and Bob implement configurable polarization measurements using a half-wave plate (HWP) and a polarization beam splitter (PBS), whose outputs are connected to single-photon detectors ($D_{+}, D_{-}$). Dashed boxes indicate the components of Alice’s and Bob’s measurement apparatus.}
    \label{fig:spdc_setup}
\end{figure}

A typical photonic Bell experiment employs a nonlinear optical process known as spontaneous parametric down-conversion (SPDC) to generate pairs of polarization-entangled photons. The generic setup is illustrated in Fig.~\ref{fig:spdc_setup}~\cite{ramanathan2018steering,smania2020experimental,seguinard2023experimental,piveteau2024optimization}. A pump laser illuminates a nonlinear crystal, such as a $\beta$-barium borate (BBO) crystal, producing photon pairs at lower energy. These photons are emitted into two distinct spatial modes directed toward two distant measurement stations, conventionally labeled Alice and Bob.

To ensure high-quality entanglement, any residual spatial, spectral, or temporal distinguishability between the photon pairs is minimized. This is typically achieved by coupling the photons into single-mode fibers (SMFs) to define well-controlled spatial modes, followed by narrow-band interference filters (F) to reduce spectral correlations. Temporal compensation elements may also be included to balance group-velocity walk-off in the nonlinear medium. These filtering steps increase the fidelity of the photon pair but reduce the overall collection efficiency.

After the filtering stage, each photon travels to a measurement station. Both Alice and Bob implement configurable polarization measurements using a half-wave plate (HWP) followed by a polarization beam splitter (PBS). The PBS separates the photon into one of two possible outputs, which are then detected by single-photon detectors, here labeled $D_{+}$ and $D_{-}$. The detection events from all output channels are time-tagged and analyzed in coincidence logic units with a narrow coincidence window to identify true photon pairs.

A key experimental parameter is the pump laser power, which directly influences the photon pair production rate. At low pump powers, the rate of multi-pair emissions is negligible, leading to high-visibility entangled correlations. However, as the pump power increases, higher-order SPDC processes generate multiple pairs within the same detection window. This results in accidental coincidences that degrade the observed Bell violation. Understanding and modeling this trade-off between higher event rates and increased accidental counts is essential for optimizing quantum technologies that rely on Bell inequality violations, such as quantum random number generation and device-independent quantum key distribution. A systematic analysis of these effects will be developed in the subsequent sections.

\section{Role of accidental counts}
\label{sec:accidental_counts}

We investigate quantum entanglement generated via SPDC, using a pulsed laser with frequency $f = 80$ MHz. Each laser pulse contains approximately $10^{10}$ photons, depending on the pump laser power, and the likelihood of an individual photon producing an entangled pair is around $10^{-12}$.

Let us account for accidental coincidences. They refer to the case when two or more entangled pairs were created in a given laser pulse, and at least one photon from different pairs was detected. Since the photons arrived at the same time bin, it is not possible for the detectors to distinguish them from an entangled pair, but since they come from two different pairs, any correlations between them are accidental.

Let $p(k, \lambda)$ denote the Poisson distribution:
\begin{equation}
	\label{eq:Poisson}
	p(k, \lambda) = \frac{\lambda^k e^{-\lambda}}{k!}
\end{equation}
representing the probability of generating $k$ entangled photon pairs in a given pulse. $\lambda$ is the average number of generated entangled pairs per single laser pulse, and it depends on the pump laser power.

Let $\eta$ denote the efficiency of the photon detectors. Let us consider the case when $k$ pairs of entangled photons were emitted in a single laser pulse. The following probabilities describe various click events observed in the experiment.

The \textit{single-click probability} is the probability that $k$ entangled photons yield exactly one detection event across both detectors (e.g., due to imperfect efficiency or loss). It is equal:
\begin{equation}
	P_{\text{single}}(\eta, k) = 4 \cdot (1 - \eta)^k \sum_{j=1}^k \binom{k}{j} \eta^j (1 - \eta)^{k-j} \left(\frac{1}{2}\right)^j.
\end{equation}
To obtain this formula, let us note that there are two parties involved, Alice and Bob, each with two detectors, giving a total factor of 4 for all possible single-click scenarios. The term $(1 - \eta)^k$ represents the probability that one of the parties (say, Bob) does not detect any of the $k$ photons arriving at their detectors. At the other side (Alice), we consider the possibilities of detecting $j$ out of $k$ photons. The term $\binom{k}{j} \eta^j (1 - \eta)^{k-j}$ gives the probability that exactly $j$ photons are detected at one side, while the remaining $k-j$ are lost. Finally, each of these $j$ detected photons has a $1/2$ chance of going to either detector, so the probability that all $j$ go to the same detector is $\left(\frac{1}{2}\right)^j$. The summation accumulates this over all valid $j$ from 1 to $k$.

The probability that at least one detector at Alice and one at Bob both click, referred to as the \textit{double-click probability}, and is given by:
\begin{equation}
	P_{\text{double}}(\eta, k) = \left(1 - (1 - \eta)^k\right)^2.
\end{equation}
Let us first consider just one of the two parties, e.g., Alice. The expression $(1 - \eta)^k$ is the probability that Alice detects none of the $k$ incoming photons. Therefore, the complementary event, in which Alice detects at least one photon, is given by $1 - (1 - \eta)^k$. Since the two parties (Alice and Bob) are independent, squaring this expression yields the joint probability that both parties detect at least one photon.

The probability that exactly one entangled photon pair is detected, one at each side, and no extra photons contribute to the coincidence, called the \textit{entangled-pair coincidence probability}, is:
\begin{equation}
	P_{\text{ent}}(\eta, k) = \binom{k}{1} \cdot \eta^2 \cdot (1 - \eta)^{2(k - 1)}.
\end{equation}
This formula captures the ideal case where exactly one photon is detected at Alice’s side and exactly one at Bob’s side, and those two photons originate from the same entangled pair. From $k$ entangled photon pairs, there are $\binom{k}{1} = k$ possible ways to choose which one pair is the one being detected. The probability that both photons from this pair are detected is $\eta^2$. All remaining $k - 1$ entangled pairs must remain completely undetected, meaning that each of their two photons must be lost; the probability for this is $(1 - \eta)^{2(k - 1)}$.

Note that:
\begin{equation}
	\label{eq:PdoublePentK1}
	P_{\text{double}}(\eta, 1) = P_{\text{ent}}(\eta, 1) = \eta^2.
\end{equation}

The expected rate of events associated with a detection probability function $P(\eta, k)$ is expressed as the average:
\begin{equation}
	r(\eta, \lambda, P) = \mathbb{E}_{k \sim \text{Poisson}(\lambda)}\left[ P(\eta, k) \right] = \sum_{k = 1}^{\infty} p(k, \lambda) \cdot P(\eta, k)
\end{equation}
This functional expresses that we are averaging the probability $P(\eta, k)$ over the Poisson-distributed values of $k$.

The total number of observed single and double (or more) clicks in time interval $t$ are then given by:
\begin{subequations}
	\begin{equation}
		\label{eq:s}
		s(\eta, \lambda, t) = f \cdot t \cdot r(\eta, \lambda, P_{\text{single}}),
	\end{equation}
	\begin{equation}
		\label{eq:c}
		c(\eta, \lambda, t) = f \cdot t \cdot r(\eta, \lambda, P_{\text{double}}).
	\end{equation}
\end{subequations}

To describe imperfections in quantum state preparation and measurement, we model the quantum state using Werner noise. In this model, the state is denoted by $\rho_v$, where $v = 1$ corresponds to a pure state $\proj{\Phi}$ and $v = 0$ corresponds to the maximally mixed state $\frac{1}{4} \openone$. The Werner state is then a convex combination:
\begin{equation}
	\label{eq:Werner}
	\rho_v = v \proj{\Phi} + (1 - v)\frac{1}{4} \openone.
\end{equation}
Importantly, the state $\ket{\Phi}$ need not be the same as the maximally entangled state $\ket{\phi^{+}}$.
We assume that the entangled-pair coincidences account to the proper preparation of the quantum states, and that the remaining double-clicks accounts to the white noise. The measured visibility in the Werner noise model is thus given as:
\begin{equation}
	\label{eq:v_model}
	v(\eta, \lambda) = \frac{r(\eta, \lambda, P_{\text{ent}})}{r(\eta, \lambda, P_{\text{double}})},
\end{equation}
where $r(\eta, \lambda, P_{\text{ent}})$ denotes the expected rate of true entangled coincidences, and $r(\eta, \lambda, P_{\text{double}})$ is a normalization term related to the total number of double clicks, including accidental ones.

\section{Noise model for accidental counts}
\label{sec:noise_model}

Denote by $c_{obs}$ and $s_{obs}$ the observed number of double and single clicks, respectively. In the first order approximation, we have
\begin{equation}
	\frac{s_{obs}}{c_{obs}} \approx \frac{P_{\text{single}}(\eta, 1)}{P_{\text{double}}(\eta, 1)} \approx \frac{2 \cdot (1 - \eta)}{\eta},
\end{equation}
and thus
\begin{equation}
	\label{eq:eta_approx}
	\eta \approx \frac{2}{2 + s_{obs} / c_{obs}}.
\end{equation}

The theoretical model for visibility given by~\eqref{eq:v_model} in the second order is approximated for $\lambda \approx 0$ as
\begin{equation}
	\begin{aligned}
		v &\approx \frac{p(1,\lambda) P_{\text{ent}}(\eta, 1) + p(2,\lambda) P_{\text{ent}}(\eta, 2)}{p(1,\lambda) P_{\text{double}}(\eta, 1) + p(2,\lambda) P_{\text{double}}(\eta, 2)} \\
		&= \frac{P_{\text{ent}}(\eta, 1) + \frac{\lambda}{2} P_{\text{ent}}(\eta, 2)}{P_{\text{double}}(\eta, 1) + \frac{\lambda}{2} P_{\text{double}}(\eta, 2)} \approx \frac{1}{P_{\text{double}}(\eta, 1)} \times \\
		&\left( P_{\text{ent}}(\eta, 1) + \frac{\lambda}{2} P_{\text{ent}}(\eta, 2) \right) \times \left( 1 - \frac{\lambda}{2} \frac{P_{\text{double}}(\eta, 2)}{P_{\text{double}}(\eta, 1)} \right) \\
		&\approx \frac{P_{\text{ent}}(\eta, 1)}{P_{\text{double}}(\eta, 1)}   + \\
		&\qquad + \frac{\lambda}{2} \left( \frac{P_{\text{ent}}(\eta, 2)}{P_{\text{double}}(\eta, 1)} - P_{\text{ent}}(\eta, 1) \frac{P_{\text{double}}(\eta, 2)}{P_{\text{double}}(\eta, 1)^2} \right).
	\end{aligned}
\end{equation}
From this equation, using~\eqref{eq:PdoublePentK1}, we get
\begin{equation}
	\label{eq:v_approx}
	v \approx 1 - \frac{\lambda}{2} \times \xi(\eta),
\end{equation}
where
\begin{equation}
	\label{eq:xi}
	\xi(\eta) \equiv \frac{P_{\text{double}}(\eta, 2) - P_{\text{ent}}(\eta, 2)}{\eta^2}.
\end{equation}

We denote by $\mathbb{B}$ the Bell operator corresponding to a particular Bell inequality, defined as a linear combination of measurement observables such that for the ideal state $\proj{\Psi}$,
\begin{equation}
	T = \Tr(\mathbb{B} \proj{\Psi}),
\end{equation}
where $T$ is the Tsirelson bound for the Bell expression, and $\ket{\Psi}$ achieves this bound.

To incorporate measurement imperfections, we assume they introduce a fixed error. That is, even if the ideal state $\proj{\Psi}$ is prepared, due to these imperfections the observed Bell value becomes
\begin{equation}
	\Tr(\mathbb{B} \proj{\Psi}) - \beta = T - \beta,
\end{equation}
where $\beta$ quantifies the effect of measurement imperfections.

Now, suppose the prepared quantum state is not $\proj{\Psi}$ but some other state $\proj{\Phi}$, possibly due to imperfections in state preparation, and also measurement misalignment. The noiseless Bell value is then
\begin{equation}
	\label{eq:B_noiseless}
	\Tr(\mathbb{B} \proj{\Phi}) - \beta = \alpha \Tr(\mathbb{B} \proj{\Psi}) - \beta = \alpha T - \beta,
\end{equation}
for some $\alpha \in [0,1)$ representing the degradation due to state preparation and measurement imperfections. Clearly, the Tsirelson bound is not reached in such a scenario.

Let $\phi(\lambda)$ denote the actually created state which includes noise effects arising from accidental counts, as described earlier. For $\lambda \to 0$, we recover the state being prepared by the optical configuration, i.e.,
\begin{equation}
	\label{eq:phi_lambda}
	\phi(\lambda) \to \proj{\Phi}.
\end{equation}
From the Werner noise~\eqref{eq:Werner} and the approximate visibility model~\eqref{eq:v_approx}, it follows that for small $\lambda$ and fixed $\xi = \xi(\eta)$,
\begin{equation}
	\phi(\lambda) \approx \bar{\phi}(\lambda) \equiv \left(1 - \frac{\lambda}{2} \cdot \xi \right)\proj{\Phi} + \frac{1}{4} \cdot \frac{\lambda}{2} \cdot \xi \openone.
\end{equation}
We now compute the observed Bell value for this noisy state:
\begin{equation}
	\begin{aligned}
		\Tr(\mathbb{B} \bar{\phi}(\lambda)) - \beta &= \Tr\left(\mathbb{B} \left(1 - \frac{\lambda \xi}{2} \right) \proj{\Phi} + \mathbb{B} \cdot \frac{\lambda \xi}{8} \openone\right) - \beta \\
		&= \left(1 - \frac{\lambda \xi}{2} \right) \Tr(\mathbb{B} \proj{\Phi}) + \frac{\lambda \xi}{8} \Tr(\mathbb{B}) - \beta.
	\end{aligned}
\end{equation}
Since the Bell operators considered here are based on correlators, they all satisfy $\Tr(\mathbb{B}) = 0$, the second term vanishes, and we are left with
\begin{equation}
	\label{eq:BtildePhi}
	\Tr(\mathbb{B} \bar{\phi}(\lambda)) - \beta = \alpha T \left(1 - \frac{\lambda \xi}{2} \right) - \beta = -\frac{\alpha T \xi}{2} \lambda + (\alpha T - \beta).
\end{equation}

This justifies the following linear approximation $\bar{B}(\lambda)$ for the Bell value, $B = B(\eta,\lambda)$ for fixed $\eta$ and $0 \leq \lambda \ll 1$ accounting for both systematic imperfections of measurement and state preparation:
\begin{equation}
	\label{eq:B_linear_fit}
	B(\lambda) \approx \bar{B}(\lambda) \equiv a \cdot \lambda + b.
\end{equation}
Comparing~\eqref{eq:BtildePhi} with the linear fit model from~\eqref{eq:B_linear_fit}, we identify the slope and intercept as:
\begin{equation}
	\begin{aligned}
		a &= -\frac{\alpha T \xi}{2}, b = \alpha T - \beta \text{, or} \\
		\alpha &= -\frac{2 a}{T \xi}, \beta = -\frac{2 a}{\xi} - b.
	\end{aligned}
\end{equation}
These expressions link the linear regression coefficients to physical parameters of the system and measurement noise.
Finally, the observed value of the Bell expression is:
\begin{equation}
	\label{eq:B_extrapolation}
	B(\eta,\lambda) = \alpha T v(\eta,\lambda) - \beta.
\end{equation}
Note that $v$, given in~\eqref{eq:v_model}, is a non-linear function of $\eta$ and $\lambda$.

\section{Predicting Bell value}
\label{sec:predictions}

In this section, we apply the noise model introduced in Section~\ref{sec:noise_model} to predict the value of Bell expressions as a function of the laser pump strength. To do so, we analyze experimental data reported in~\cite{piveteau2024optimization}, where Bell violations were measured for varying pump powers in a photonic experiment based on SPDC. The dataset provides, for each pump setting, the observed Bell value along with relevant experimental parameters such as the number of detected events, heralding rate, and coincidence count rates. By fitting our model to this data, we determine parameters characterizing the contribution of accidental counts and use them to extrapolate Bell values at untested pump levels. This enables us to quantify the trade-off between source brightness and the quality of Bell violations.

\begin{table}[htbp]
	\begin{tabular}{c|c|c}
		\hline
		$i$ & events per second & Bell value \\ \hline
		1   & 70000             & 2.6502     \\
		2   & 52000             & 2.6760     \\
		3   & 36000             & 2.7026     \\
		4   & 20000             & 2.7150     \\
		5   & 12000             & 2.7369     \\
		6   & 8000              & 2.7443     \\
		7   & 4000              & 2.7609     \\ \hline
	\end{tabular}
	\caption{$i$ is the experiment number, where additional data are provided in Tab.~\ref{tab:prop2_counts}.}
	\label{tab:prop2_Bell}
\end{table}

Consider the observed values of CHSH from the experiment, $B_{obs}^{(i)}$, provided in Tab.~\ref{tab:prop2_Bell}.
Tab~\ref{tab:prop2_counts} presents raw data collected during the experiment. Here, $c_{obs}^{(i)}$ and $s_{obs}^{(i)}$ are the observed number of double and single clicks, and $t^{(i)}$ is the measurement time (in seconds) for $i$-th experimental case.  Using the formula~\eqref{eq:eta_approx} and data from Tab.~\ref{tab:prop2_counts}, we obtained $\eta \approx 0.1134$.

\begin{table}
	\centering
	\begin{tabular}{|r||r|r|r||r|r|}
		\hline
		$i$ & $c_{obs}^{(i)}$ & $s_{obs}^{(i)}$ & $t^{(i)}$ & $\lambda_{calc}^{(i)}$ & $B_{calc}^{(i)}$ \\ \hline
		1   & 37892989        & 549605351       & 540       & $0.0649$               & $2.6486$         \\
		2   & 43322946        & 660223194       & 832       & $0.0488$               & $2.6760$         \\
		3   & 40639747        & 646698789       & 1112      & $0.0346$               & $2.6999$         \\
		4   & 40767056        & 631786125       & 2000      & $0.0195$               & $2.7255$         \\
		5   & 41386494        & 668668812       & 3332      & $0.0120$               & $2.7382$         \\
		6   & 40381162        & 650423503       & 5000      & $0.0078$               & $2.7453$         \\
		7   & 36888729        & 590756887       & 10000     & $0.0036$               & $2.7524$         \\ \hline
	\end{tabular}
	\caption{Observed counts and visibility at various time intervals for CHSH. $c_{obs}^{(i)}$ and $s_{obs}^{(i)}$ are the single and double clicks, and $t^{(i)}$ are times of collecting events. $\lambda_{calc}^{(i)}$ is the calculated average numbers of generated entangled pairs per single laser pulse, and $B_{calc}^{(i)}$ are the linear fit of Bell values using~\eqref{eq:B_linear_fit}. The experiment number correspond to cases from Tab.~\ref{tab:prop2_Bell}.}
	\label{tab:prop2_counts}
\end{table}

The values of the average number of generated entangled pairs per single laser pulse, $\lambda$, in each case for $i$-th row, were determined as the value minimizing the difference between $c(\eta, \lambda, t)$ defined in~\eqref{eq:c} and $c_{obs}^{(i)}$. The results are denoted as $\lambda_{calc}^{(i)}$.

A linear fit was then performed between the calculated $\lambda_{calc}^{(i)}$ and the measured Bell values $B_{obs}^{(i)}$.

The calculated fit parameters for~\eqref{eq:B_linear_fit} are:
\begin{equation}
	\label{eq:linear_fit_a_b_values}
	a = -1.6917, \quad b = 2.7585
\end{equation}
and a root-mean-square error (RMSE) of:
\begin{equation}
	\text{RMSE} = 0.0053.
\end{equation}
This negative slope demonstrates that increasing pump power (and thus $\lambda$) leads to a reduction in visibility due to the growing influence of multi-pair events. The value of $b$ suggests that if there were no accidental counts, the state prepared $\ket{\Phi}$, cf.~\eqref{eq:phi_lambda}, would yeald CHSH value $2.7585 < T \approx 2.8284$. Careful optimization is required to balance generation rate with entanglement quality for quantum applications. The obtained values of visibilities for linear fit, $v_{calc}^{(i)}$, are given in Tab.~\ref{tab:prop2_counts}. We use~\eqref{eq:B_extrapolation} to extrapolate the Bell value, denoted $B_{calc}$, and estimate events per second with~\eqref{eq:c}.

\begin{table}[]
	\begin{tabular}{c|c|c}
		\hline
		events per second & $B_{calc}$ & $\lambda_{calc}$ \\ \hline
		93240   & 2.625 & 0.0849 \\
		113636  & 2.6   & 0.1022 \\
		207254  & 2.5   & 0.1769 \\
		322581  & 2.4   & 0.2614 \\
		470588  & 2.3   & 0.3576 \\
		655738  & 2.2   & 0.4684 \\
		888889  & 2.1   & 0.5972 \\
		1212121 & 2     & 0.7490 \\  \hline
	\end{tabular}
	\caption{Extrapolated values of asymptotic generation rates when increasing the number of events per second using CHSH as the certificate. The maximal asymptotic generation rate is for the value of CHSH equal $2.3$. $B_{calc}$ is obtained from~\eqref{eq:B_extrapolation} using the value~\eqref{eq:linear_fit_a_b_values} calculated from the experimental data from Tabs~\ref{tab:prop2_Bell} and~\ref{tab:prop2_counts}.}
	\label{tab:chsh_extrapolation}
\end{table}

\section{Conclusions}
\label{sec:conclusions}

In this work, we investigated how accidental counts, arising from increased laser pump power, affect the violation of Bell inequalities in photonic experiments. We focused on scenarios relevant to quantum random number generation (QRNG) and quantum key distribution (QKD), where both high event rates and strong Bell violations are crucial.

We proposed a noise model that incorporates the effect of accidental detections into the observed quantum state via a simple admixture of white noise, following the Werner state formalism. This model allowed us to analytically express how the Bell value degrades with increasing accidental count probability, which grows with the source brightness. We introduced a method to predict the observed Bell value based on experimental parameters, notably the laser pump power and measured detection statistics.

Using experimental data from~\cite{piveteau2024optimization}, we applied our model to two specific Bell expression, viz. the CHSH inequality. Our approach yielded predictions of Bell values beyond the tested range of pump powers, showing good agreement with experimental trends. This demonstrates the model's utility in optimizing Bell-based quantum protocols under realistic laboratory conditions.

Our results emphasize the importance of balancing source brightness and noise when designing practical QRNG and QKD implementations. By enabling performance prediction across experimental regimes, our framework supports the design and calibration of quantum optical setups that aim to achieve high rates without compromising security certifications derived from Bell violations.

Future work could extend the model to include other sources of imperfections, such as detector jitter or multi-pair emissions, and test its applicability across a broader range of quantum platforms and Bell expressions.

\section*{Acknowledgements}

This work was supported by the Knut and Alice Wallenberg Foundation through the Wallenberg Centre for Quantum Technology (WACQT). PM was supported by the European Union’s Horizon Europe research and innovation programme under grant agreement No 101080086/NeQST.
We are grateful to the Anonymous Reviewer of~\cite{piveteau2024optimization} for pointing out the research problem raised in this work.

\end{document}